\documentclass[jap,preprint,preprintnumbers,amsmath,amssymb,nobibnotes,showkeys,aip]{revtex4}

\usepackage{graphicx}% Include \vec{figure} files
\usepackage{dcolumn}% Align table columns on decimal point
\usepackage{bm}% bold math

\begin{document}                % INITIALIZE - DONT CHANGE % %  %

\title{Electrostatic force spectroscopy of near surface localized states}

\author{Aykutlu D\^{a}na}
\affiliation{Advanced Research Labs, Department of Physics, Bilkent
University, 06800 Ankara, Turkey}

\author{Yoshihisa Yamamoto}
\affiliation{Edward L. Ginzton Laboratory, Stanford University, Stanford, California 94305-4085} %

\begin{abstract}
Electrostatic force microscopy at cryogenic temperatures is used
to probe the electrostatic interaction of a conductive atomic
force microscopy tip and electronic charges trapped in localized
states in an insulating layer on a semiconductor. Measurement of
the frequency shift of the cantilever as a function of tip-sample
shows discrete peaks at certain voltages when the tip is located
near trap centers. These discrete changes in frequency is
attributed to one by one filling of individual electronic states
when the quantized energies traverses the substrate conduction
band fermi energy as tip-sample voltage is increased. Theoretical
analysis of the experiment suggests that such measurement of the
cantilever frequency shift as a function of bias voltage can be
interpreted as an AC force measurement, from which spectroscopic
information about the location, energy and tunneling times of
localized states can be deduced. Experimental results from study
of a sample with InAs quantum dots as trap centers is presented.

\end{abstract}
\maketitle
\section{Introduction}

As the semiconductor device size continues to shrink, new methods
for characterization of electrical properties of materials and novel
devices on the nanometer scale are required
\cite{semiconductorroadmap}. Detection of impurities,
characterization of complex material stacks and interfacial
properties, non-destructive electrical characterization of
ultra-thin gate and capacitor dielectrics and 3D dopant profiling
are few of the challenges faced as the device size decreases to
nanometer scale. The challenge of electrical characterization of
novel devices with smaller number of atoms motivates development of
a technique that provides qualitative information about individual
electronic states available within the devices.

Since its introduction \cite{afmquate}, the atomic force microscope
(AFM) and its spin-off techniques have been widely used in imaging
and characterization of semiconductor surfaces. Electrostatic force
based imaging techniques\cite{ssrmreview} such as Kelvin probe
microscopy (KPM), scanning capacitance microscopy (SCM) and scanning
spreading resistance microscopy (SSRM) among others have been used
to electrically characterize surfaces. Still, an in-situ,
non-destructive technique for characterization of semiconductor
surfaces and sub surface structures at the single electronic state
level is not available. Because of its high force sensitivity, AFM
has been used to detect the presence of individual electronic
charges on the sample surface or inside layers near the surface
\cite{singleelectron1,singleelectron2,singleelectron3}. However, to
be able to use the AFM to characterize individual states, we still
need to develop a method of obtaining information about the
location, energy and dynamics of states on or near the semiconductor
surface through force measurements.

In this article, to address this problem, we present a technique
based on measurement of electrostatic forces between a conducting
AFM tip and charges localized at near-surface electronic states. A
conducting AFM tip is used both as a gate electrode and as an
electrometer that senses accumulated charge on the sample.
Measurement of electrostatic forces between the tip and the sample
as a function of the tip-sample bias voltage provide information
about the location, energy and tunneling dynamics of localized
states. Regarding this measurement technique as a spectroscopy, we
refer to it as electrostatic force spectroscopy (EFS) from here
on. In the following sections, we begin by formulating the
problem, defining the sample structure to which this technique
applies, give a theoretical analysis of the EFS experiment, and
provide experimental results.

\section{Theoretical analysis of electrostatic force spectroscopy of localized states}

When biased a conducting AFM tip is brought near a conducting
sample surface, due to the finite tip sample capacitance, charges
of opposite sign accumulate in the tip and on the sample surface.
This electrostatic interaction can be measured through deflection
of the cantilever or through perturbation of its resonance
frequency. If the sample is a semiconductor or a layered
semiconductor/insulator structure with localized states, the
electrostatic interaction between the tip and the sample deviates
from a simple capacitor and presence of localized states has to be
accounted for in the analysis of the electrostatic forces. Based
on a model of the sample, measurement of electrostatic interaction
as a function of tip location and tip-sample bias voltage can
provide data that can be inverted to give information about the
location and energy of localized states or doping concenterations.
Characterization of electronic states associated with traps inside
thin dielectrics, states at semiconductor interfaces, states due
to defects and presence of adsorbates are important for the
semiconductor technology. Therefore, we choose to restrict
ourselves to a metal-insulator-semiconductor configuration with
low density of localized states, as described in the following
subsection.

\subsection{Tip-Sample configuration for an EFS experiment}

The proposed sample configuration is schematically shown in Fig.
1(a). The conductive AFM tip is placed above a
insulator-on-conductor structure, with a tip-sample separation of
$z_{ts}$. In an actual experiment, the insulating layer can be a
dielectric material deposited or grown on top of the highly
conductive region, or a thin dielectric film otherwise placed on a
flat conductive sample. In the analysis presented here, the sample
is assumed to be a monolithic semiconductor where the conductive
region and the insulating dielectric layer is defined by doping. The
band diagram in such a configuration is illustrated in Figure 1b.
The localized states can be due to impurities, dislocations,
interface traps or intentionally introduced states due to presence
of quantum dots. The sample structure presented here has certain
benefits. The localized states are inside an insulating layer so
charge trapped in these states are not screened by free carriers.
Also, since there is no doping in the top layer, the 3D potential
profile generated by the tip is simple to analyze analytically.
Moreover, the localized states can be charged and discharged by
tunneling of carriers from the bulk through the insulator. This
modulation of the charge and resulting perturbation of the
electrostatic force forms the basis of the proposed detection
method.

\subsection{Electrostatic model for calculating local potentials}
Analysis of the EFS scheme begins with a model that describes the
electrostatic force between the tip and sample and the potential
profile inside the insulating layer. The electrostatic problem
described by the tip-sample system can be analyzed analytically
through a piecewise model of the tip. The charge density on the
tip surface and the potential profile in the insulating region can
be calculated approximately by modeling the tip as the union of a
conic section and a spherical section as shown in Fig. 2. The
overall tip-sample capacitance is assumed to be the sum of
individual dihedral capacitances\cite{dihedral} formed by
infinitesimal surface elements on the tip (shown as location A in
Fig. 2) and corresponding surface elements on the surface
(location B in Fig. 2).

The sphere-cone model of the tip can be used to estimate the local
potential $V(x,z)$ (see Fig. 2) inside the insulating layer. The
calculation of $V(x,z)$ can be done, by noting that coordinate x
is related to the geometrical model variable $\varphi$ by a single
valued function $g(\varphi)$ as
\begin{equation}\label{xvsphi}
x=g(\varphi)=r\sin\varphi
+\frac{1-\cos\varphi}{\sin\varphi}\times[z_{ts}+r(1-\cos\varphi)].
\end{equation}

The local potential $V(x,z)$ is then given by
\begin{equation}\label{vofxz}
V(x,z)=\frac{V_{ts}z}{\epsilon_r}\times[\frac{\varphi[z_{ts}+r(1-\cos{\varphi})]}{\sin\varphi}+\frac{d_{ins}}{\epsilon_r}]^{-1}
\end{equation}
where $\varphi=g^{-1}(x)$. Eq. (\ref{vofxz}) agrees with a finite
element analysis solution of the potential within $\%5$ if
$\rm{d_{ins}/\epsilon_r \ll z_{ts}}$ and r, $\rm{z_{ts}\lesssim
r}$ and $\rm{|x|\lesssim 4r}$.

For a flat metal sample, the electrostatic force estimated through
this model (sphere-cone model) can be expressed in terms of the
tip length $H_{tip}$, tip radius r, tip-sample separation
$z_{ts}$, and tip half-cone angle $\theta_0$ as the sum of conical
and spherical contributions
\begin{equation}\label{dihedralforcetotal}
F_{sc}=F_{sphere}+F_{cone}
\end{equation}
where the spherical and conical terms are given by
\begin{equation}\label{dihedralsphere}
F_{sphere}=V_{ts}^2\pi\epsilon_0r^2\frac{1-\sin{\theta_0}}{z_{ts}[z_{ts}+r(1-\sin{\theta_0})]}
\end{equation}
\begin{equation}\label{dihedralcone}
F_{cone}=\frac{V_{ts}^2\pi\epsilon_0\sin{\theta_0}^2}{(\pi/2-\theta_0)^2}\times
[\ln{\frac{H_{tip}}{z_{ts}+r(1-\sin{\theta_0})}}-1+\frac{r\tan{\theta_0}}{z_{ts}+r(1-\sin{\theta_0})}].
\end{equation}

The validity of this model can be tested through measurements of
force gradients of a biased tip as a function of the tip sample
separation. It is seen from the data presented in Fig. 3 that by
fitting the tip radius only, the model given in Eq.
(\ref{dihedralforcetotal}) predicts the tip-sample capacitance
qualitatively with less than $\%5$ error in the range $r/2\lesssim
z_{ts}\lesssim 4r$.

\subsection{Model for charging of the localized states}

For a given tip-sample geometry, and a given bias voltage
$V_{ts}$, the energy of a localized state i, away from the tip
axis a distance x and at a height $\rm{h_i}$ from the ground plane
(see Fig. 1 and Fig. 2) is given by
\begin{equation}\label{eiwithlocalpotential}
  E_i=E_{i,0}-eV(x,h_i)
\end{equation}
where e is the electronic charge, $\rm{V(x,h_i)}$ is given by Eq.
(\ref{vofxz}) and $\rm{E_{i,0}}$ is the energy of the state under
zero bias. For a given sample, if we define the dimensionless
parameter $\rm{\alpha(x,z_{ts})}$ as
\begin{equation}\label{alphaxzts}
  \alpha(x,z_{ts})=\frac{z+d_{ins}/\epsilon_r}{\varphi(z_{ts}+r(1-cos\varphi))/sin\varphi + d_{ins}/\epsilon_r}
\end{equation}
where $\rm{\varphi}$ is related to x through Eq. (\ref{xvsphi}),
we can rewrite Eq. (\ref{eiwithlocalpotential}) as

\begin{equation}\label{chargingequationwithalpha}
E_i=E_{i,0}-\frac{eV_{ts}h_i}{z_{ts}\epsilon_r+d_{ins}}\alpha(x,z_{ts}).
\end{equation}

It worth noting that, for states on the tip axis,
$\rm{\alpha(0,z_{ts})=1}$ and Eq. (\ref{chargingequationwithalpha})
reduces to a simple voltage divider.

 In thermal equilibrium, charge $\rm{q_i}$ of state i can
be calculated through thermal statistics as

\begin{equation}\label{equilibriumcharge}
q_i=-\frac{e}{1+\exp[(E_i-E_f)/k_BT]}
\end{equation}
where $\rm{k_BT}$ is the thermal energy.

When $E_i$ is modulated in time, if the tunneling time
$\Gamma_i^{-1}$ is finite but does not strongly depend on
$\rm{V_{ts}}$, the time dependent charge $\rm{\tilde{q_i}}$ can be
calculated through a first order differential equation as
\begin{equation}\label{timedependentqi}
  \Gamma_i^{-1}\frac{d\tilde{q_i}}{dt}=-\tilde{q_i}+q_i(t).
\end{equation}
Here $\rm{q_i(t)}$ denotes $q_i$ calculated through Eq.
(\ref{equilibriumcharge}), and the time dependece is due to
modulation of $\rm{V_{ts}}$ or $\rm{z_{ts}}$. $\rm{\Gamma_i}$
stands for the tunneling rate for state i for the given DC bias
condition. The approximation presented in Eq.
(\ref{timedependentqi}) would be valid only for a small signal
modulation of the charge, since $\rm{\Gamma_i}$ depends
exponentially on the potential barrier and can not be assumed
constant over a large modulation of the barrier. If a small signal
sinusoidal modulation of $V_{ts}$ or $z_{ts}$ with frequency
$\omega$ is present, $\tilde{q_i}$ will be given by a sinusiod
that has a phase $\phi$ that depends on the modulation frequency
and tunneling rate $\Gamma_i$ as
\begin{equation}\label{phaseqi}
  \phi=-\arctan(\omega/\Gamma_i).
\end{equation}
The modulated charge amplitude $\rm{\tilde{q_i}}$ can be
calculated through
\begin{equation}\label{modulatedchargeamplitude}
  \tilde{q_i}=\langle\frac{\partial q_i}{\partial V_{ts}}\rangle\tilde{V}_{ts}
+\langle\frac{\partial q_i}{\partial z_{ts}}\rangle\tilde{z}_{ts}
\end{equation}
where the derivatives are calculated through Eqs. (\ref{vofxz}),
(\ref{eiwithlocalpotential}) and (\ref{equilibriumcharge}),
averages denoted by brackets are taken over the modulation ranges
of respective modulated variables. Here $\rm{\tilde{V}_{ts}}$ and
$\rm{\tilde{z}_{ts}}$ are the modulation amplitudes of bias and
tip-sample separation respectively.

\subsection{Electrostatic force model in the presence of localized states}

The electrostatic interaction of the tip and the ground plane can
be analyzed through the sphere-cone model accurately. In the
presence of localized states with charges $\rm{q_i}$, there is
additional contribution to the force from individual charges. For
the sake of simplicity, the electrostatic force $\rm{F_e}$ that
includes contributions from the localized charges and the
background will be approximated by a parallel plate capacitor
model given by \cite{singleelectron3}
\begin{equation}\label{parallelplatemodel}
F_e\cong\frac{\epsilon_r^2\xi}{(\epsilon_rz_{ts}+d_{ins})^2}\times[\frac{\pi
r^2\epsilon_0V_{ts}^2}{2}+\sum_i{\frac{2h_iq_iV_{ts}}{\epsilon_r}}]
\end{equation}
where $\rm{\xi}$ is a geometric correction factor that can be
calculated by equating $\rm{F_e}$ of Eq. \ref{parallelplatemodel}
with all $q_i$ being identically zero, to the electrostatic force of
Eq. (\ref{dihedralforcetotal}). In the parameter range
$r/4<z_{ts}<2r$, $\xi$ varies from 0.9 to 4.2 being equal to 1 if
$z_{ts}/r = 0.4$. It is important to note that Eq.
(\ref{parallelplatemodel}) is written assuming that the interaction
is due to localized charges on the tip axis and a lumped charge due
to tip-sample capacitance concenterated at the tip apex. In reality,
force due to each localized state has to be corrected by integrating
the force between $\rm{q_i}$ and the charge distribution on the tip.
Also, extension of Eq. (\ref{parallelplatemodel}) to include effect
of charges away from the tip axis can be done by including effect of
geometry. The simplification made in derivation of the force in Eq.
(\ref{parallelplatemodel}) assuming a lumped parallel plate
capacitor model, will have effect only on the magnitude of the
forces from individual charges.

\subsection{Modulation of electrostatic force: localized state signatures}

When an AC modulation of the tip-sample separation or tip-sample
bias voltage is present, Eq. (\ref{parallelplatemodel}) can be
used to estimate the AC modulated electrostatic force. Since the
objective of EFS experiment proposed in this work is to extract
information about localized states through measurement of forces,
in this subsection we will analyze the contribution from the
localized states only. The AC force due to localized states can be
calculated through
\begin{equation}\label{ACforcefromlocalizedstates}
  \tilde{F}_e=\sum_i \frac{\partial F_e}{\partial q_i}\tilde{q}_i
\end{equation}
where $\rm{F_e}$ and $\rm{\tilde{q}_i}$ are given by Eqs.
(\ref{parallelplatemodel}) and (\ref{timedependentqi})
respectively. Eq. (\ref{ACforcefromlocalizedstates}) includes only
the contribution due to modulation of the charges in the localized
states and does not account for the modulated background force due
to presence of the bulk of the sample. The background contribution
can be calculated by direct differentiation of Eq.
(\ref{parallelplatemodel}) with respect to $\rm{z_{ts}}$ or
$\rm{V_{ts}}$ with $q_i$ set to zero. This background contribution
will be analyzed in the following subsections, since it proves to
be a significant effect in the detection process.

Each term in the sum on the right-hand side of Eq.
(\ref{ACforcefromlocalizedstates}) contains information about the
corresponding localized state, and we shall refer to it as the
\textit{signature} of that particular state. The signature force
is a function of $V_{ts}$, the tip location with respect to the
sample  $z_{ts}$, the energy of the state $E_{i,0}$ and its height
from the ground plane $h_i$. Therefore, measuring the modulated
force for a set of values of $V_{ts}$ and $z_{ts}$ we can estimate
$E_{i,0}$ and $h_i$.

When only a modulation of the bias voltage $\tilde{V}_{ts}$ is
present, and tip location is fixed $\tilde{z}_{ts}=0$, the
signature for state i is
\begin{equation}\label{signaturewithbiasmodulation}
  \tilde{F}_{e,i}=\frac{2\epsilon_r h_i \xi V_{ts}}{(\epsilon_r z_{ts} + d_{ins})^2}\langle\frac{\partial q_i}{\partial
  V_{ts}}\rangle\tilde{V}_{ts}.
\end{equation}
Conversely, when tip-sample separation is modulated only and
$\tilde{V}_{ts}=0$, the signature is
\begin{equation}\label{signaturewithztsmodulation}
  \tilde{F}_{e,i}=\frac{2\epsilon_r h_i \xi V_{ts}}{(\epsilon_r z_{ts} + d_{ins})^2}\langle\frac{\partial q_i}{\partial
  z_{ts}}\rangle\tilde{z}_{ts}.
\end{equation}

The dependence of the signatures in Eqs.
(\ref{signaturewithbiasmodulation}) and
(\ref{signaturewithztsmodulation}) on $V_{ts}$ and $z_{ts}$ is
presented in Figs. 4 and 5, for a set of typical experimental
parameters. It is seen from Fig. 4 that, each state appears as a
distinct peak when we plot $\tilde{F}_e$ against $V_{ts}$. This
can be inituitively understood noting that, as the bias voltage is
increased, the energy of the state traverses the fermi energy of
the ground plane and it is charged. Only when the state energy is
close to the fermi energy, the state charge can be modulated by a
modulation of the local potential. This modulation amplitude has
the energy dependence of the derivative of the thermal
distribution and thus the AC force amplitude appears as a peak
when plotted versus $\rm{V_{ts}}$. The signature voltage $V_{s,i}$
at which the force has peak amplitude is given through Eq.
(\ref{chargingequationwithalpha}) for a state a distance x away
from the tip axis as
\begin{equation}\label{signaturevoltage}
V_{s,i}=\frac{E_{i,0}(\epsilon_r
z_{ts}+d_{ins})}{eh_i\alpha(x,z_{ts})}
\end{equation}
and in the limit of infinitesimal modulation amplitude, the width
$\Delta V_{s,i}$ of the peak in terms of bias voltage is
\begin{equation}\label{signaturevoltagewidth}
\Delta V_{s,i}=\frac{2k_BT(\epsilon_r
z_{ts}+d_{ins})}{eh_i\alpha(x,z_{ts})}.
\end{equation}

 It is noted from Fig. 4 that, as the temperature is decreased
and $k_BT$ becomes small compared to the modulation of $E_i$, the
averaging of the derivative of $q_i$ (denoted by the brackets in
Eqs. (\ref{signaturewithbiasmodulation}) and
(\ref{signaturewithztsmodulation})) over the modulation range
causes the signature to deviate from a gaussian-like peak, and Eq.
(\ref{signaturevoltagewidth}) no longer applies.

If $\rm{\Delta V_{s,i}}$ can be measured accurately, then we can
estimate $\rm{E_{i,0}}$ from Eqs. (\ref{signaturevoltage}) and
(\ref{signaturevoltagewidth}) as
\begin{equation}\label{eiestimation}
E_{i,0}=\frac{2k_BTV_{s,i}}{\Delta V_{s,i}}.
\end{equation}

To reduce the error in estimation of state parameters, one can
repeat the EFS measurement changing only the tip location. From a
set of EFS data taken at different values of the $\rm{z_{ts}}$, it
is possible to determine $\rm{V_{s,i}}$, $\rm{\Delta V_{s,i}}$ and
$\rm{\partial V_{s,i}/\partial z_{ts}}$. These parameters can then
be used to solve for the three unknowns x, $\rm{E_{i,0}}$ and
$\rm{h_i}$ through Eqs. (\ref{signaturevoltage}),
(\ref{signaturevoltagewidth}) and (\ref{eiestimation}) uniquely.

In a case where measurement of $\rm{\Delta V_{s,i}}$ has large
error bounds due to imperfections of the measurement setup,
another method has to be devised to extract location, height and
energy of the state. Due to the cylindrical symmetry of the tip,
the potential of Eq. (\ref{vofxz}) will have circular
equipotential contours. If the tip is scanned in the x-y plane
keeping $V_{ts}$ and $z_{ts}$ constant, and $\rm{V_{s,i}}$ is
plotted as a function of x and y, resulting image will exhibit
circular patterns whose radii can be related to the experimental
parameters and parameters of the state i using Eq.
(\ref{signaturevoltage}). The data resulting from such a
measurement can also be used to estimate $h_i$ as will be
illustrated in the experimental sections.

\subsection{Measurement of electrostatic forces: Self-oscillation technique}

The electrostatic force, modulated or DC, causes a deflection of
the cantilever which can then be detected through a secondary
detector, such as a laser interferometer. The minimum detectable
electrostatic force is given by the themomechanical noise limit,
regardless of measurement frequency or technique. However,
modulation frequency or measurement technique can be important in
optimization of the signal-to-noise ratio (SNR), since secondary
detector can not be assumed noiseless. For example, a typical
laser interferometer used for cantilever deflection detection in
our experiments, has a noise floor of $\rm{2\times10^{-3}}$
$\rm{\AA/\sqrt{Hz}}$.  Referring to the figure 4 and 5, modulated
electrostatic forces due to single states are on the order of
$\rm{10^{-12}}$ Nt for a typical experimental configuration. If a
cantilever with a spring constant of say $\rm{k_0}$=1 Nt/m and
quality factor Q $\sim$ $\rm{10^4}$ is used, the peak deflection
amplitude for a state signature will be on the order of
$\rm{10^{-12}}$ m if modulation frequency is near DC and
$\rm{10^{-8}}$ m if modulation frequency $\rm{\omega}$ is on
resonance with the cantilever mechanical resonance
$\rm{\omega_0}$. The secondary detection limited charge
sensitivity can be estimated to be 0.1 e/$\rm{\sqrt{Hz}}$ near DC
and $\rm{10^{-5}}$ e/$\rm{\sqrt{Hz}}$ on resonance. However,
thermomechanical noise floor for our cantilevers is
$\rm{4\times10^{-16}}$ Nt/$\rm{\sqrt{Hz}}$ at 4 K independent of
$\rm{\omega}$, and it corresponds to a fundamental limit for
charge resolution of $\rm{4\times10^{-4}}$ e/$\rm{\sqrt{Hz}}$.
Thermomechanical noise is dominant in the overall force
measurement if $\rm{\omega\simeq\omega_0}$.

Modulation frequency and technique is also important in
realization of the EFS experiment. In order for the analysis
presented for the modulation of $\rm{V_{ts}}$ to hold, $z_{ts}$
must be kept constant, otherwise Eq.
(\ref{signaturewithbiasmodulation}) will no longer describe the
signature force correctly. In practice, this can be done by
suppression of the cantilever oscillation by a feedback loop.
However, modulation of the bias voltage with
$\rm{\omega\simeq\omega_0}$ requires tracking of the frequency
shift of the cantilever due to the z-gradient of the background
electrostatic force which is given by
\begin{equation}\label{backgroundfrequencyshift}
\Delta\omega=-\frac{\omega_0\xi\pi\epsilon_r^3\epsilon_0
  r^2V_{ts}^2}{2k_0(\epsilon_rz_{ts}+d_{ins})^3}
\end{equation}
where $\rm{k_0}$ is the spring constant of the cantilever.

The difficulties one has to overcome in order to realize the EFS
experiment by modulating $\rm{V_{ts}}$ can be solved if
$\rm{z_{ts}}$ is modulated instead of $\rm{V_{ts}}$. Modulation of
$\rm{z_{ts}}$ has the two benefits: First, there is no need
actively to suppress modulation of $\rm{V_{ts}}$ to validate
assumptions made in analysis, since it can be biased by an
external DC voltage source. Second, if the cantilever is
oscillated by positive feedback or a phase-locked loop system on
its resonance, the modulation of $\rm{z_{ts}}$ will automatically
be always on resonance with the cantilever. These benefits
motivate the use of self-oscillation of the cantilever.

Technical description of self-oscillation feedback can be found
elsewhere\cite{frequencyshiftmethod1,frequencyshiftmethod2}.
Self-oscillation technique was generally used to detect the force
gradients due to time invariant interactions. This method can be
applied to measurement of AC forces through frequency shift
measurements. The method uses feedback to sustain the oscillation
of the cantilever on its resonance, by measuring the AC deflection
$\tilde{z}_{ts}$, phase shifting by $\rm{\pi/2}$, conditioning it
for amplitude control and feeding it back as a drive force
$\tilde{F}_D$. The effect of the external feedback can be written
by setting $\rm{\tilde{z}_{ts}(t)=\tilde{z}_{ts}\sin(\omega t)}$
and $\rm{\tilde{F}_{D}(t)=\tilde{F}_{D}\cos(\omega t)}$. When an
external signal force $\rm{\tilde{F}_s(t)=\tilde{F}_s\sin{(\omega
t+\phi)} }$ is present, the oscillation amplitude
$\rm{\tilde{z}_{ts}}$ and oscillation frequency
$\rm{\delta\omega}$ can be calculated through

\begin{equation}\label{oscillationamplitudethroughforce}
\tilde{z}_{ts}\cong\frac{Q}{k_0}(\tilde{F}_D+\tilde{F}_s\sin{\phi})
\end{equation}
and
\begin{equation}\label{frequencyshiftthroughforce}
\delta\omega\cong\frac{\omega_0}{2k_0\tilde{z}_{ts}}\tilde{F}_s\cos{\phi}
\end{equation}
where Q is the quality factor of the cantilever. Approximations
presented in Eqs. (\ref{oscillationamplitudethroughforce}) and
(\ref{frequencyshiftthroughforce}) can be assumed valid if
$\rm{\delta\omega\ll\omega_0}$.

Inserting $\rm{\tilde{F}_s=\tilde{F}_{e,i}}$ from Eq.
(\ref{signaturewithztsmodulation}), the signature of a state can
be measured in the frequency shift of the cantilever in
self-oscillation configuration as

\begin{equation}\label{freqshiftsignature}
  \delta\omega_i=\frac{\omega_0\epsilon_r h_i \xi V_{ts}\cos{\phi}}{k_0(\epsilon_r z_{ts} + d_{ins})^2}\langle\frac{\partial q_i}{\partial
  z_{ts}}\rangle.
\end{equation}

The effect of temperature and oscillation amplitude on the overall
SNR for this measurement scheme is illustrated in Fig. 6. The
phase $\rm{\phi}$ can be estimated by measuring
$\rm{\tilde{z}_{ts}}$ and $\rm{\delta\omega_i}$ for a single
state. Tunneling rate $\rm{\Gamma_i}$ for the state can then be
related to $\rm{\phi}$ through Eq. (\ref{phaseqi}).

In the self-oscillation method based measurement of the
signatures, the total frequency shift is the sum of the background
frequency shift of Eq. (\ref{backgroundfrequencyshift}) and
signature frequency shifts given by Eq. (\ref{freqshiftsignature})
as

\begin{equation}\label{frequencyshiftspectroscopy}
  \Delta\omega_{efs}=\Delta\omega+\sum_{i}\delta\omega_i
\end{equation}

The minimum detectable charge in the frequency shift method is
again given by the thermomechanical detection limit although
method of detection is through measurement of frequency shift
instead of deflection. Also, presence of the self-oscillation
feedback does not affect the value of minimum detectable force.
Only difference is, force noise translates to a fundamental
frequency noise
\section{Experiment}

The EFS experiments presented here uses a home built low
temperature AFM system that can operate down to 4.2 K. A fiber
interferometer serves as the secondary detector. The laser
wavelength is $\rm{\lambda}$=1310 nm, with 100 $\rm{\mu W}$
optical power incident on the cantilever, and measured noise floor
for deflection detection is $\rm{2\times10^{-3}}$
$\rm{\AA/\sqrt{Hz}}$. Commercial Pt/Ir coated cantilevers with
spring constants of $\rm{k_0}$=2.8 Nt/m and resonant frequencies
of $\rm{\omega_0}$=75 KHz are used. Supplier specified tip lengths
are $\rm{H_{tip}}\simeq10$ $\rm{\mu m}$ and the half-cone angle of
the tip is $\rm{20^0}$. The tip radius is not specified but can be
extracted through force measurements to be $\rm{r\simeq}$ 20 nm.
The quality factor of the cantilevers Q is around 150 in air,
15,000 at room temperature in vacuum, and range from 30,000 to
45,000 as temperature is decreased from 77.3 K to 4.2 K.
Mechanical actuation of the cantilever oscillation using a
piezoelectric element can produce spurious freqeuncy shifts
because mechanical structures can have multiple resonances near
the operation frequency. Therefore, an electrostatic actuation
scheme is used to oscillate the cantilever because of constant
phase and amplitude response in the frequency range of interest.

The sample is chosen to contain InAs QDs embedded in insulating
GaAs since similar samples have been previously extensively
studied for characterization of QD energy levels by optical and
electrical methods\cite{petroff1,petroff2}. Based on previous
capacitance spectroscopy experiments \cite{petroff3} incorporating
similar InAs QDs, we expect the QD energies to be from 250 meV to
100 meV below the GaAs conduction band edge. It is also estimated
that the number of confined energy levels and values of confined
energies depend on QD size and up to 12 confined energy levels are
estimated as the QD base diameter approaches 40 nm. Growth
conditions have a strong effect on QD energy levels
\cite{petroff1,inasleveltuning} since gallium can replace indium
in the dots and this alloying affects QD band gap. Although it is
not possible to know the quantized energies of QDs only knowing
the growth conditions, a rough estimation of the energy levels is
still important for choosing the right experimental parameters of
tip-sample separation and bias voltage range.

The sample is a molecular beam epitaxy (MBE) grown GaAs structure.
First, a GaAs buffer layer with silicon doping of density
$10^{18}$ $\rm{cm^{-3}}$ and thickness of 500 nm is grown,
followed by an undoped GaAs layer of 15 nm thickness. Then a
monolayer of InAs wetting layer was grown followed by a single
layer of InAs QDs. The dots were capped by a undoped GaAs capping
layer of 15 nm thickness. Topographical AFM image of a test sample
grown under same conditions without capping layer, the QDs were
found to be about 20 nm in diameter, about 4 nm tall, with a
surface density of $\rm{10^{10}}$ $\rm{cm^{-2}}$.

%Also, The maximum electric field before breakdown
%$\rm{E_{max}}\sim5\times10^{8}$ V/m, provides an upper bound for
%the bias voltage.

Contact mode topographic images of the surface was obtained prior
to the EFS experiment to ensure the flatness and cleanliness of
the surface. Force-distance curve with $\rm{V_{ts}}$=0V provides
information about the location of the surface, $\rm{z_s}$. The
drift of the scanner in x,y and z directions were characterized by
repeating imaging and force-distance measurements with few minutes
intervals, before and after the experiments. It was seen that when
the AFM is operated at 4 K, the drift was insignificant
($\rm{\sim}$ 2 nm) over an hour and can be ignored.

\subsection{Observation of the wetting layer}

It is known through previous experiments\cite{petroff1} that the
InAs wetting layer (WL) forms a 2 dimensional electron gas (2
DEG). In a crude approximation, it can be regarded as a collection
of localized states and should present some form of signature in
the EFS data. Study of charging of the WL in our EFS experiment is
interesting, since it produces a large signal due to large number
of electronic states. Also, the ground state energy of the WL with
respect to the GaAs conduction band edge can provide a reference
for the EFS data. Finally, the WL provides states at all locations
on the sample and we do not have to find a proper location to
observe the WL. The band gap of GaAs at room temperature is
$\rm{E_{GaAs}}$=1.52 eV at 4.2 K, and surface pinning is assumed
to be at the middle of the band gap. In previous photoluminescence
measurements of similar structures, WL optical transition occurs
1.42 eV. Therefore, if we assume for the sake of interpretation of
EFS data, that WL is a localized state, the corresponding electron
energy for that state under zero bias condition will be
$\rm{E_{wl,0}}$ = 330 meV. The EFS data shown in Fig. 7 is
collected with a tip sample separation of $\rm{z_{ts}}$=14.5 nm,
where $\rm{z_{ts}}$ is measured by a force-distance curve. A
sudden change in the frequency shift indicates presence of states
that is charged when $\rm{V_{ts}}$=5.83 V. EFS experiment is
repeated at different tip-sample separations to fit the height
$\rm{h_{wl}}$ and $\rm{E_{wl,0}}$, and we find that
$\rm{h_{wl}}$=14 nm, $\rm{E_{wl,0}}$=360 meV (shown in inset of
Fig. 7). The discrepancy of EFS results may be due to pinning of
the GaAs surface at a slightly different energy than the middle of
band gap, or due to the fact that any band-bending effects were
ignored in our model.

\subsection{Observation of localized states}

In the EFS experiments performed with the aim of identifying QD
energy levels, based on theoretical calculations and preliminary
information given by the observation of WL, choosing $\rm{z_{ts}}$
to be around 20 nm and $\rm{\tilde{z}_{ts}}$ to be less than 1 nm,
we expect to obtain a SNR greater than 10 in a 100 Hz bandwith for
single states. In the capped sample, it is not possible to locate
the dots through topographical imaging since the capping produces
a flat surface. Therefore, EFS experiments were performed on a
grid of points on a flat region of the sample.

Observation of isolated single signatures depends on presence of
isolated single states in the sample. If there are many states in
the close vicinity of the tip, it is hard to distinguish
individual peaks from a single EFS measurement. This fact is
illustrated by Fig. 8(a), where many charging signatures can be
seen between $\rm{-4.5 V<V_{ts}<-2.8 V}$. It is also seen that as
$\rm{z_{ts}}$ moves from 19 nm to 20 nm, the peaks appear at a
slightly more negative voltage range $\rm{-4.9 V<V_{ts}<-3 V}$ as
expected from Eq. (\ref{signaturevoltage}). Although this expected
behaviour of states shifting towards stronger biases can be
observed, because of large number of superimposed peaks it is not
possible to identify individual signatures unambiguously. A single
isolated state signature from an EFS measurement performed at a
different location, shown in Fig. 8(b), features a single isolated
signature. For this state, $\rm{V_{s,i}}$ also shifts towards
negative voltages as $\rm{z_{ts}}$ moves from 30 nm to 35 nm.
Since this signature is well isolated, it is possible to estimate
the energy and depth of the state. Based on Eqs.
(\ref{signaturevoltage}), (\ref{signaturevoltagewidth}) and
(\ref{eiestimation}), we can estimate this state parameters to be,
$\rm{E_{i,0}=0.105}$ eV, $\rm{h_i=}$14 nm and located x=51 nm from
the tip axis.

Fig. 9(a) is an example of EFS data with no signatures of
localized states. Slowly varying background forces due to the
presence of ground plane were fitted and subtracted to clarify
that there are no distinct peaks. Fig. 9(b) shows EFS data for
another location on the sample, with six distinct peaks in both
frequency shift and oscillation amplitude. Similar signatures can
also be observed near a QD in a sample grown exactly the same but
without a capping layer (Fig. 9(c)). In the uncapped sample the
signatures dissappear when the tip is moved away from the QD,
demonstrating that the signatures are indeed due to the QD.
Energies can be fitted to each peak. The energies estimated from
Fig. 9(b), 9(c) and energies measured through conventional
capacitance spectroscopy for similar dots in a previous
measurement\cite{petroff2} are compared in Table 1.

To further illustrate the effect of tip location on $\rm{V_{s,i}}$
one can plot the signature amplitude as a function of x and y in
the vicinity of a localized state. Three signatures appear at a
bias of -4.45 V (Fig. 10(a)) and as the voltage is increased to
-5.15 V (Fig. 10(b)), the location of the signature peak defines a
circular pattern, equivalent to an equipotential contour which is
defined by Eq. (\ref{eiwithlocalpotential}). Energy and height of
the state can be estimated as $\rm{h_i}$=14.5 nm and
$\rm{E_{i,0}}$=205 meV by fitting Eq. (\ref{eiwithlocalpotential})
to the data.

\section{Conclusions}

A simplified theory of EFS generalized to a family of samples that
has localized states inside a thin insulating layer is presented.
The technique is capable of extracting information about individual
localized states with nanometer resolution and $4\times10^{-4}$
electronic charge sensitivity. Application of the technique to InAs
quantum dots embedded in a semiinsulating GaAs matrix is presented
as a demonstration. The presented theory gives guidelines for choice
of cantilever and sample parameters for a given application of EFS.
Potential applications include, high resolution 3D dopant profiling
in semiconductors, characterization of novel thin gate dielectrics,
and nondestructive characterization of self-assembled monolayer
materials for nanoelectronic devices.

 \acknowledgments The authors would like to thank JST and Stanford University for
 their continued support during this work.

\begin{table}
\label{table1} \caption{Electron energy levels inferred from
previous capacitive measurements for 20 nm base diameter capped
dots\cite{ribeiro2}, theory for 11.3 nm base diameter capped
dots\cite{jeongnimkim} and this experiment involving 40 nm base
diameter uncapped dot. Electron energies are shifted to match the
ground state energies, $\rm{E_{s-1}}$. The calculation by Kim et al.
\cite{jeongnimkim} does not take into account Coulomb charging
effects and estimates $\rm{E_{s-1}}$ to be 231 meV below the GaAs
conduction band minimum. }
\begin{tabular}
{c c c c}
Energy level &Theory&Capacitance data &EFS for capped QD \\
\hline
$\rm{E_{s-1}}$ (meV) &0& 0 & 0 \\
$\rm{E_{s-2}}$ (meV)&  &19 & 35\\
$\rm{E_{p-1}}$ (meV) & 84&74 & 57\\
$\rm{E_{p-2}}$ (meV)&  &82 & 63\\
$\rm{E_{p-3}}$ (meV)& 111&100 & 88\\
$\rm{E_{p-4}}$ (meV)&  &110 & 93\\
\end{tabular}
\end{table}

\begin{figure}
\begin{center}
\resizebox{0.5\textwidth}{!}{\includegraphics{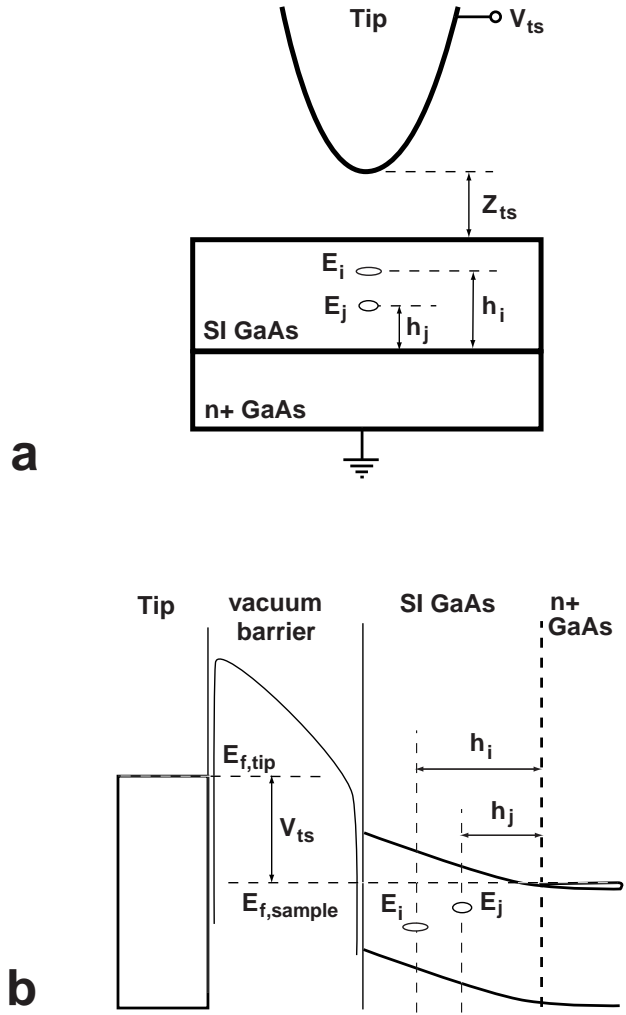}}
\end{center}
\caption{Schematic of the EFS experiment. a,  Configuration of the
tip and the sample that contains the states to be studied. States
with energies Ei at heights hi are located inside an insulating
layer on top of a highly conductive ground plane. In the analysis
and experiments presented in this work, the sample is chosen to be a
monolithic semiconductor. The insulating and conducting regions are
determined by doping. b, Illustration of the energy band diagram. }
\end{figure}

\begin{figure}
\begin{center}
\resizebox{0.65\textwidth}{!}{\includegraphics{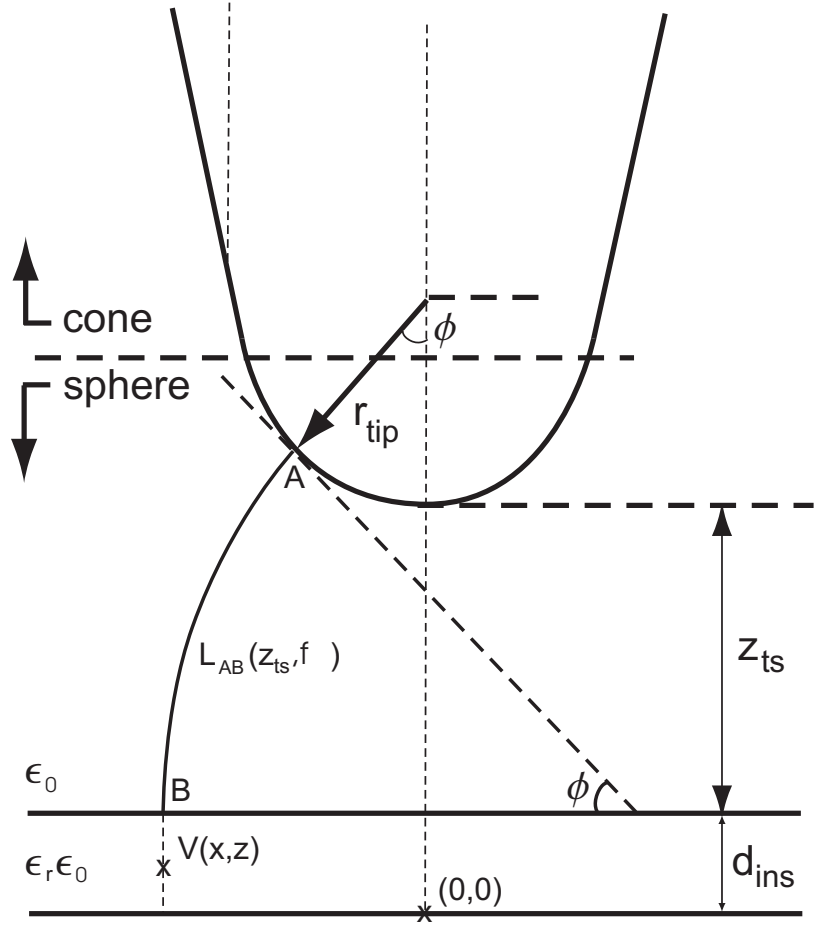}}
\end{center}
\caption{Description of the sphere-cone model of the tip-sample
interaction. The AFM tip is modelled as the union of spherical and
conical sections. The electrostatic problem is solved by integrating
contributions of individual dihedral capacitors formed between
surface elements on the tip (point A) and corresponding surface
elements on the sample (point B). Tip-sample electrostatic force and
potential profile inside the dielectric ($\rm{V(x,z)}$) can be
accurately described by the model.}
\end{figure}

\begin{figure}
\begin{center}
\resizebox{0.5\textwidth}{!}{\includegraphics{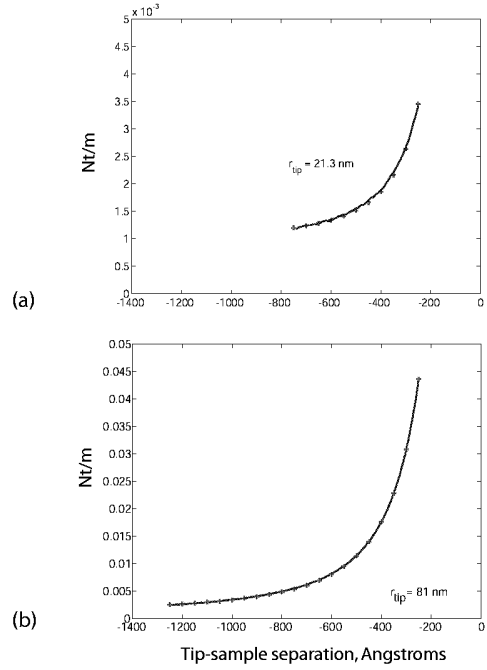}}
\end{center}
\caption{ Electrostatic force gradient $\rm{\partial F_e/\partial
z}$ measured through frequency shift of the cantilever and
theoretical estimation through Eq. \ref{dihedralforcetotal} by
fitting the tip radius. a) A fresh tip has a fitted radius of r =
21.3 nm. b) After contact imaging and deposition of metal on the
surface through pulsing of the bias voltage, the tip radius
increases to r = 81 nm. }
\end{figure}

\begin{figure}
\begin{center}
\resizebox{0.85\textwidth}{!}{\includegraphics{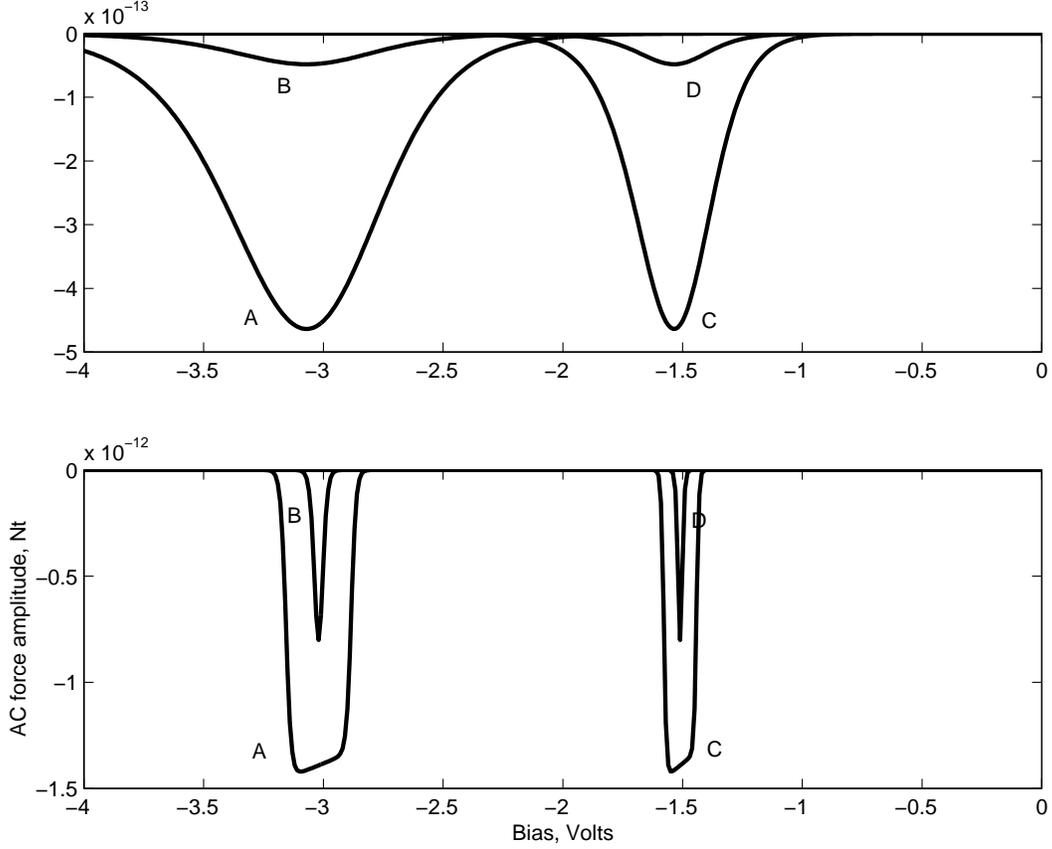}}
\end{center}
\caption{Theoretical force signatures of two states, a) under
modulation of the tip-sample separation at T=77 K. Curves A and B
are calculated for a state with the parameters $E_{i,0}$=0.1 eV,
$h_i$ =10 nm, with modulation amplitude $\tilde{z}_{ts}$ = 1 nm and
0.1 nm respectively. Curves C and D are for a state with the
parameters $E_{i,0}$ =0.1 eV, $h_i$ =20 nm. b) Same as (a) except
T=4 K. The voltage at which the force peak occurs, and the width of
the peak in terms of bias voltage provide information about energy
and location of the state. The sample is chosen to be GaAs, with
$\epsilon_r$ = 13.6. Thickness of the insulating section is
$d_{ins}$ = 30 nm. Tip radius is r=20 nm and $z_{ts}$ = 20 nm.
Negative amplitudes denote the fact that the modulated force has
opposite phase with the modulation of the tip-sample separation. }
\end{figure}

\begin{figure}
\begin{center}
\resizebox{0.85\textwidth}{!}{\includegraphics{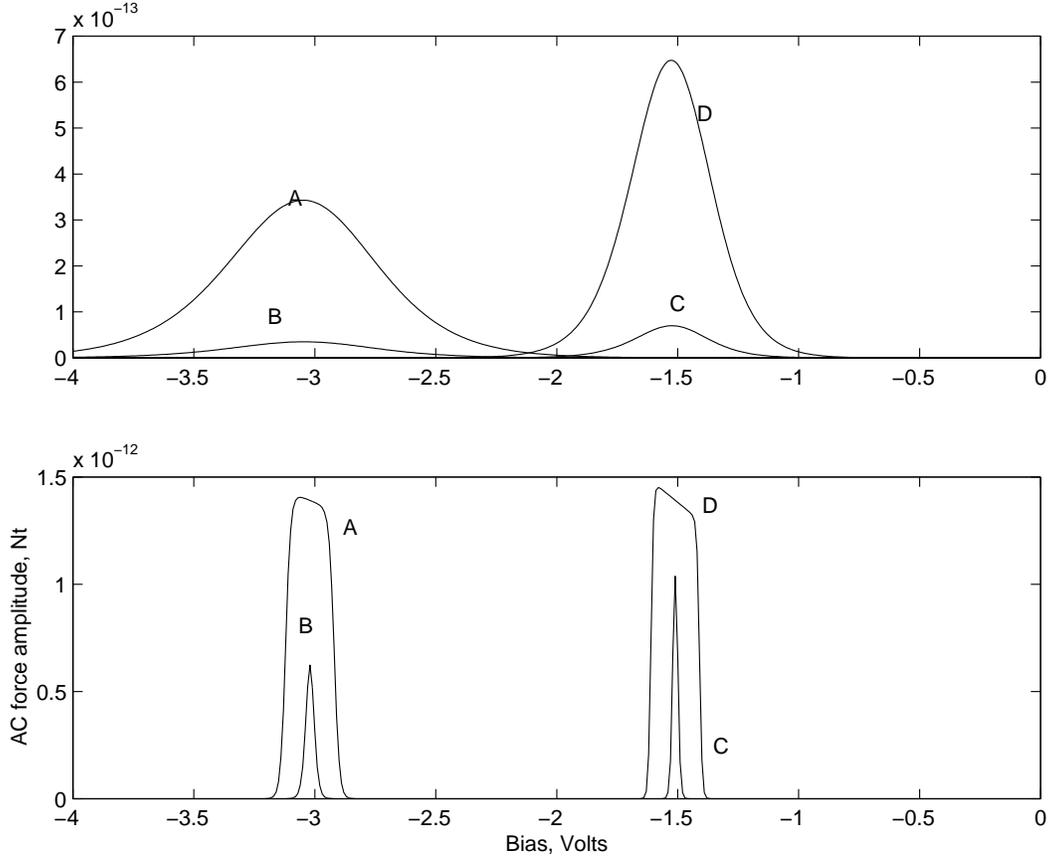}}
\end{center}
\caption{Theoretical force signatures of two states, a) under
modulation of the bias voltage $\rm{V_{ts}}$ at T=77 K. Curves A and
B are calculated for a state with the parameters $E_{i,0}$=0.1 eV,
$h_i$ =10 nm, with modulation amplitude $\tilde{V}_{ts}$ = 10 mV and
100 mV respectively. Curves C and D are for a state with the
parameters $E_{i,0}$ =0.1 eV, $h_i$ =20 nm. b) Same as (a) except
T=4 K. Sample paremeters are the same as in Figure 4. Positive
amplitudes denote the fact that the modulated force has same phase
with $\tilde{V}_{ts}$.}
\end{figure}

\begin{figure}
\label{overallsnr}
\begin{center}
\resizebox{0.75\textwidth}{!}{\includegraphics{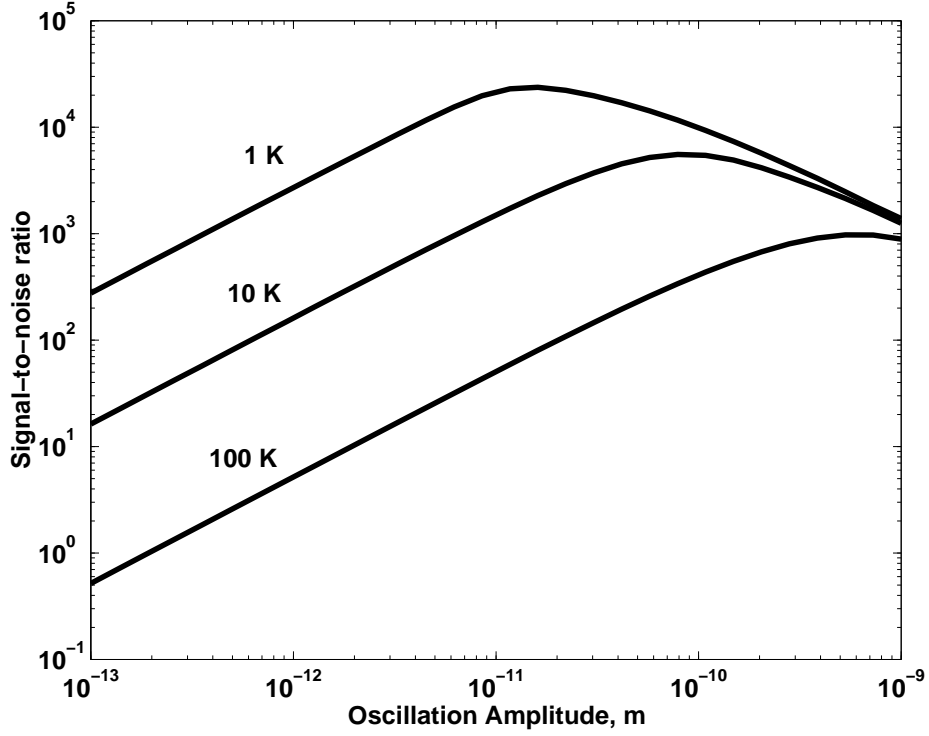}}
\end{center}
\caption{Signal-to-noise ratio for a single localized state in the
frequency measurement technique as a function of temperature and
oscillation amplitude $\rm{\tilde{z}_{ts}}$. The state parameters
are $\rm{E_{i,0}}$=350 meV, $\rm{h_i}$=14 nm. Total dielectric
thickness is $\rm{d_{ins}}$=30 nm and $\rm{\epsilon_r}$=13.6.
Cantilever resonance frequency is $\rm{\omega_0/2\pi}$=73 KHz, and
spring constant is $\rm{k_0}$=2.8 Nt/m. Tip sample separation is
$\rm{z_{ts}}$=12 nm. Frequency detection is limited by noise of the
electronics at higher oscillation amplitudes. This fact causes SNR
to decrease if the oscillation amplitude is increased above an
optimal value which is about 1 $\rm{\AA}$ at 10 K.}
\end{figure}

\begin{figure}
\begin{center}
\resizebox{0.65\textwidth}{!}{\includegraphics{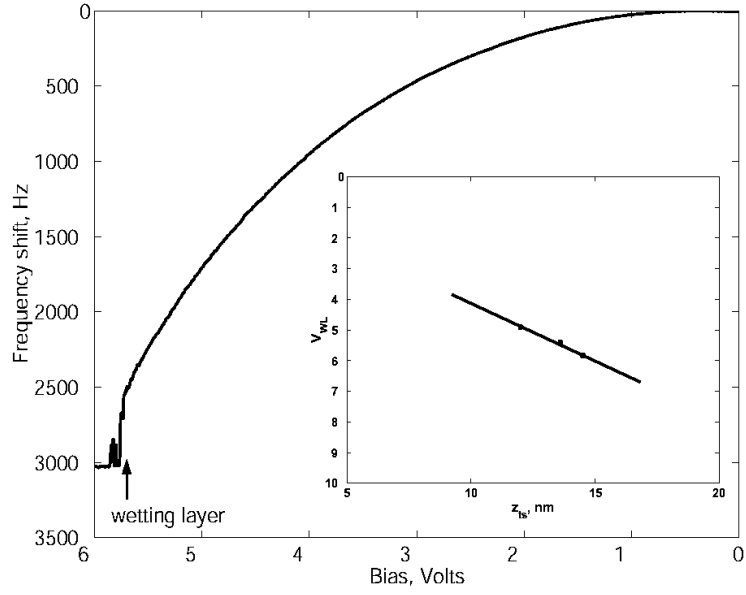}}
\end{center}
\caption{Observation of the InAs wetting layer (WL). The frequency
shift due to background electrostatic forces follows a parabola
which shows a sudden jump, indication of presence of a large number
of states. Inset shows theoretical estimation of the signature
voltage $\rm{V_{wl}}$ as a function of $\rm{z_{ts}}$. Fitting to
data, the states which cause the jumps is estimated to be
$\rm{h_{wl}}$= 14 nm above the ground plane and at an energy 25 meV
below the GaAs conduction band. }
\end{figure}

\begin{figure}
\begin{center}
\resizebox{0.65\textwidth}{!}{\includegraphics{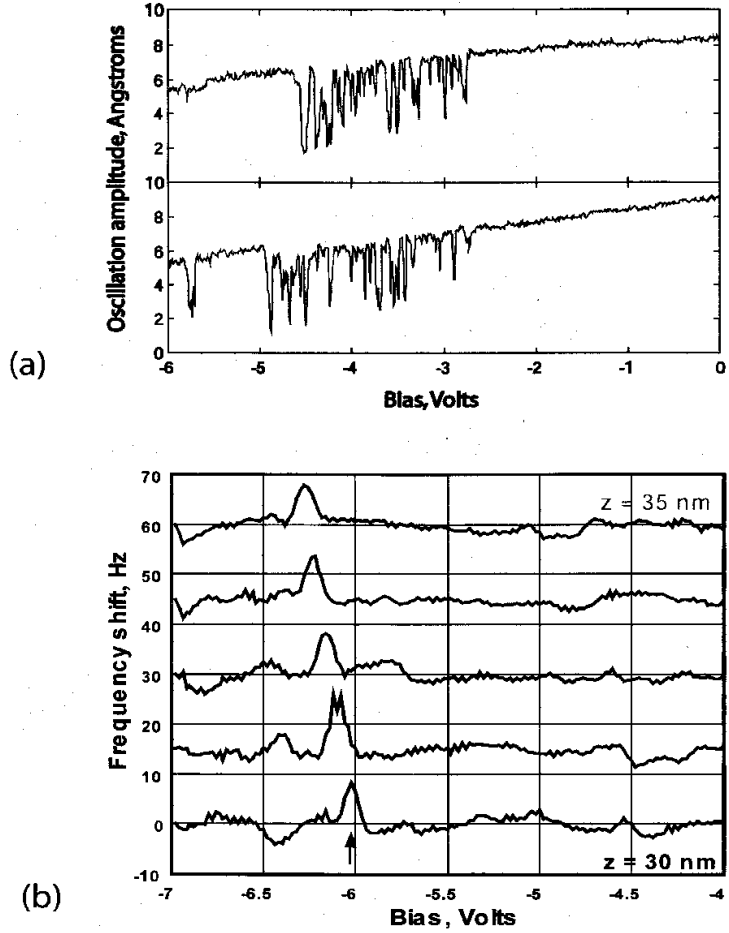}}
\end{center}
\caption{Observation of localized states. a) Multiple signature
peaks appear in the oscillation amplitude of the cantilever, in the
bias voltage range $\rm{-4.5 V<V_{ts}<-2.8 V}$ when $\rm{z_{ts}}$=19
nm (top curve). As the tip is moved away from the sample, to
$\rm{z_{ts}}$=20 nm, signature peaks move to stronger bias voltages
$\rm{-4.9 V<V_{ts}<-3.0 V}$ (bottom curve). b) A single state
signature can be isolated in the EFS data taken at a different
location of the sample. Signature voltage $\rm{V_{s,i}}$ moves to
stronger biases as the tip-sample separation $\rm{z_{ts}}$ is
increased from 30 nm to 35 nm. }
\end{figure}

\begin{figure}
\begin{center}
\resizebox{0.5\textwidth}{!}{\includegraphics{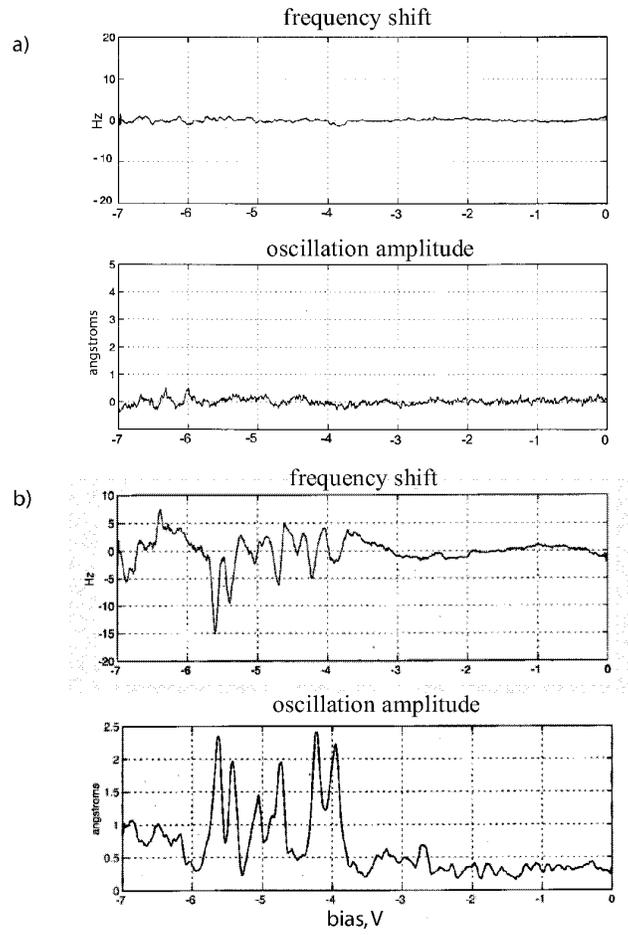}}
\end{center}
\caption{ Observation of localized states. a) Example of an EFS data
with no signatures, b) on a site where there are localized states as
evident from signatures. }
\end{figure}

\begin{figure}
\begin{center}
\resizebox{0.5\textwidth}{!}{\includegraphics{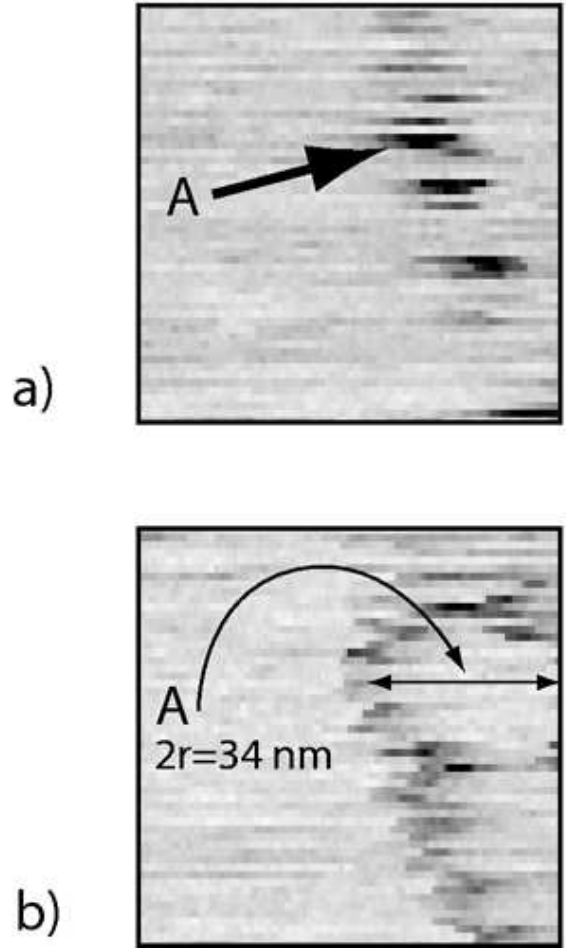}}
\end{center}
\label{amplitudevsxy} \caption{Signature amplitude plotted as a
function of x-y position of the tip in the vicinity of localized
states. Tip height is $\rm{z_{ts}}$=20 nm and the bias voltage is
a)$\rm{V_{ts}}$=-4.45 V, b)$\rm{V_{ts}}$=-5.15 V. The signature
located at point A first appears at $\rm{V_{ts}}$=-4.45 V and has a
radius of 17.3 nm at $\rm{V_{ts}}$=-5.15 V. Theoretical estimate for
the state, $\rm{h_i}$=14.5 nm and $\rm{E_{i,0}}$=205 meV correctly
estimates the appearance and evolution of the signature.}
\end{figure}

\end{document}